\documentclass[prb,aps,twocolumn]{revtex4}
\usepackage{amsmath,bm,graphicx,epsfig,citesort}

      \input{amssym.def}
      \input{amssym}
      \newcommand {\mm}[1] {\ifmmode{#1}\else{\mbox{\(#1\)}}\fi}
      \newcommand{\real} {\mm{{\Bbb R}}}

\newcommand{\utwi}[1]{\mbox{\boldmath $ #1$}}
\newcommand{\beq}{\begin{equation}}
\newcommand{\eeq}{\end{equation}}
\newcommand{\bea}{\begin{eqnarray}}
\newcommand{\eea}{\end{eqnarray}}
\newcommand{\bml}{\begin{mathletters}}
\newcommand{\eml}{\end{mathletters}}

\newcommand{\bx}{{\utwi{x}}}

\begin{document}

\title{Importance of chirality and reduced flexibility of protein side chains: \\
A  study with square and tetrahedral lattice models
}

\author{Jinfeng Zhang$^1$,  Yu Chen$^2$, Rong Chen$^{1,2,3}$, and
Jie Liang$^{1}$ \footnote{ Corresponding author.  Phone: (312)355--1789,
fax: (312)996--5921, email: {\tt jliang@uic.edu} }
}
\affiliation{
$^1$Departments of Bioengineering and
$^2$Information \& Decision Science\\
 University of Illinois at Chicago \\
845 S.\ Morgan St, Chicago, IL 60607
\\
\vspace*{.1in}
$^3$Department of Business Statistics \& Econometrics\\
Peking University\\
Beijing, P.R. China
\vspace*{.1in}
\\
{\rm (Accepted by {\it J.\ Chem.\ Phys.})}
}

\date{ \today}

\begin{abstract}
{ Side chains of amino acid residues are the determining factor that
distinguish proteins from other unstable chain polymers.  In simple
models they are often represented implicitly ({\it e.g.}, by
spin-states) or simplified as one atom.  Here we study side chain
effects using two-dimensional square lattice and three-dimensional
tetrahedral lattice models, with explicitly constructed side chains
formed by two atoms of different chirality and flexibility. We
distinguish effects due to chirality and effects due to side chain
flexibilities, since residues in proteins are $L$-residues, and their
side chains adopt different rotameric states. For short chains, we
enumerate exhaustively all possible conformations.  For long chains,
we sample effectively rare events such as compact conformations and
obtain complete pictures of ensemble properties of conformations of
these models at all compactness region.  This is made possible by
using sequential Monte Carlo techniques based on chain growth method.
Our results show that both chirality and reduced side chain
flexibility lower the folding entropy significantly for globally
compact conformations, suggesting that they are important properties
of residues to ensure fast folding and stable native structure.  This
corresponds well with our finding that natural amino acid residues
have reduced effective flexibility, as evidenced by statistical
analysis of rotamer libraries and side chain rotatable bonds.  We
further develop a method calculating the exact side-chain entropy for
a given back bone structure.  We show that simple rotamer counting
underestimates side chain entropy significantly for both extended and
near maximally compact conformations.  We find that side chain entropy
does not always correlate well with main chain packing.  With explicit
side chains, extended backbones does not have the largest side chain
entropy.  Among compact backbones with maximum side chain entropy,
helical structures emerges as the dominating configurations.  Our
results suggest that side chain entropy may be an important factor
contributing to the formation of alpha helices for compact
conformations.\\
}

\end{abstract}

\maketitle
\newpage
\narrowtext

\section{Introduction}
Side chains of amino acid residues are the determining factor that
distinguish proteins from other unstable chain polymers.  Their
arrangement along primary sequence dictates the native structure of
proteins.  Side chains are also responsible for much of the complexity
of protein structure
\cite{BrombergDill94_PS,MirankerDobson96_COSB,WynnHarkinsRichardsFox96_PS,MitchellLaskowskiThornton97_P,KlimovThirumalai98_FD}. They
pack tightly, but also leave space to form voids and pockets
\cite{RichardsLim94_QRB,LiangDill01_BJ,Bagci02_JCP}.  The effects of
simplified side chain were studied in details for two dimensional
square lattice and three dimensional cubic lattice models in reference
\cite{BrombergDill94_PS}.  Such studies of simplified models played
important roles in elucidating the principles of protein folding
\cite{DillBrombergYue95_PS}, because these models allow enumeration of
all feasible conformations and calculation of exact entropy for short
chain molecules. They are also amenable to detailed sampling for
longer chain models.  However, the effects of side chains are still
not fully understood.  Several studies on side chain effects rely on
implicit models or assign different spin states to each monomer to
mimic the internal degrees of freedom of side chains
\cite{Kussell02_PRL,Micheletti98_PRL}.  It is not clear how realistic
these model are without explicit side chains.  In studies where side
chains are modeled explicitly, they are simplified: only one atom is
attached to the main chain monomer \cite{BrombergDill94_PS}.  Since
there is no internal degree of freedom for side chains of one atom,
$\chi$-angles and rotamirc states of side chains
\cite{DunbrackKarplus94_NSB} cannot be studied.

In this study, we introduce more realistic side chain models. We make
the distinction of two different side chain effects that have not been
investigated previously.  We first study the chirality effects.
Chirality effects at $C_\alpha$ atom of a residue arises because the
four atoms bonded with $C_\alpha$ are different
\cite{ElielWilen}. Side chain atom $C_\beta$ can be attached to
different positions of $C_\alpha$ relative to other atoms ($C$, $N$,
and $H$ atoms).  In Nature, all amino acid residues with side chains
are of the $L$ configuration instead of the $D$ configuration, {\it
i.e.}, the position of $C_\beta$ in relationship to $C$ and $N$ atoms
are all in a unique chiral state.  The origin of this bias is unclear
and remains a puzzle in the studies of the origin of life
\cite{Cronin88,CroninPizzarello83_ASR}.  We also study the flexibility
effects.  Flexibility effects arise because additional atoms beyond
$C_\beta$ can rotate around a single side chain chemical bond,
regardless of the chiral state of $C_\alpha$
\cite{Richards97_CMLS,Dunbrack02_COSB}.  These two effects are
different: There is a large energetic barrier for change of chiral
state, which often involves the breaking of a chemical bond. In
contrast, rotation along a single bond is relatively easy.

We use lattice models to study the effects of both chirality and
flexibility.  We introduce chirality models for two-dimensional square
lattice and three dimensional tetrahedral lattice polymers.  To model
side chain flexibility, we use explicitly side chains consisting of
two atoms, which enable the modeling of rotational degree of freedom
of side chains.  Because this leads to significant increases of the
size of conformational space, it is difficult to characterize
accurately ensemble properties of compact conformations of polymers.
We use the techniques of sequential Monte Carlo importance sampling
and resampling to generate properly weighted samples of rare events,
such as long chain conformations with maximum compactness.

We examine the distribution of all geometrically feasible
conformations of self-avoiding walks on lattice with side chains of
different chirality and flexibility.  We focus on their packing
properties and their conformational entropy.  Folding into a well
defined native structure is accompanied with large reduction in
conformational entropy.  We explore how entropy of folding is affected
by chirality and flexibility, and how it relates to the compactness of
chain polymers with side chains.  Because the absolute number of
compact conformation changes dramatically after incorporation of
chirality and side chain flexibility, it is not obvious whether these
factors helps or hinders protein folding.  Our results indicate that
chiral molecules have lower entropy of folding than achiral
models. Models with less side chain flexibility also have significant
lower entropy of folding than models with more flexible side chain.

Side chain entropy is important for protein folding and its estimation
is the subject of several studies
\cite{KussellShimada01_JMB,BrombergDill94_PS,BradySharp97_COSB,DoigSternberg95_PS,ColeWarwicker02_PS}.
To calculate side chain entropy precisely for our model polymers, we
introduce an algorithm for counting the exact number of side chain
conformations.  It is based on the observation of disconnected sets in
the conflict graph of side chain correlations.  In comparison, we find
rotamer counting \cite{PickettSternberg93_JMB} significantly
overestimates side chain entropy, and the difference is more
pronounced in most extended as well as in protein-like near compact
regions of main chain structures.

In addition, we revisit two models of protein packing, namely, the
jigsaw puzzle model and the nuts and bolts model.  We show that
packing of chain polymers with chiral side chains included is more
like nuts and bolts than jigsaw puzzles.

The results presented here are in agreement with the chiral nature of
L-amino acid residues found in natural proteins and an analysis of
flexibility of residues in real proteins.  They suggest that both
chirality and restriction in flexibility make important contributions
to protein folding.  Our presentation is organized as follows: We
first introduce side chain models for chirality and flexibility
effects in two-dimensional square lattice and three-dimensional
tetrahedral lattice.  This is followed by a description of the
parameters used in our study and the algorithm for counting side chain
conformations.  The results of chirality and flexibility on
conformational entropy by enumeration and by sequential Monte Carlo
sampling are then presented.  We then compare rotamer counting and the
exact method developed here for calculating side chain conformational
entropy.  We conclude with remarks and discussion.

\section{Models and methods}

\paragraph{Lattice side chain models.}

\begin{figure}[htb]
\vspace{0.3in}
\centerline{\epsfig{figure=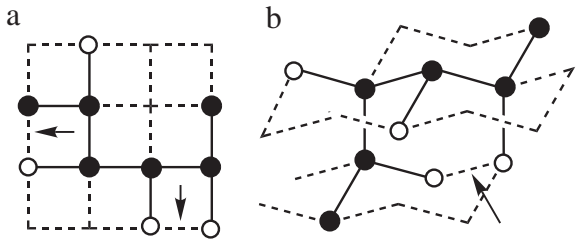,width=3in}}
\caption{\sf Lattice side chain models of 6-mers
(a) on square lattice with side chain of size one and (b) on
tetrahedral lattice with side chain of size one. Filled circles
represent main chain monomers and empty circles represent side chain
atoms. In these examples, side chains have one atom. Arrows pointing
to spatial contacts between non-bonded atoms.}
\label{Fig:LatModel} 
\end{figure}

For two-dimensional square lattice and three-dimensional tetrahedral
lattice models, a side chain consists of one or two {\it atoms\/}
attached to each main chain {\it monomer}.  There are no side chains
for the two terminal monomers following reference \cite{BrombergDill94_PS}
(Figure~\ref{Fig:LatModel}).  For three-dimensional models, we use
tetrahedral lattice instead of cubic lattice. The coordination
and bond connection of a tetrahedral unit are very similar to carbon atoms
with four chemical bonds, which is the most abundant element in
proteins.  Both chirality and flexibility can be modeled effectively
using tetrahedral lattice.  In addition, tetrahedral lattice has the
advantage that real protein structures can be well
approximated \cite{ParkLevitt95_JMB,XiaHuangLevittSamudrala00_JMB}.

\paragraph{Models for chirality.}
A molecule that is distinct from its mirror image is a chiral
molecule.  The idea of ``chirality'' in molecule goes back to Pasteur,
who observed in 1848 that crystals of tartaric acid rotated polarized
light in different directions, either to the right (D for ``{\it
dextro}'') or left (L for ``{\it levo}'') \cite{Flapan2000}.  Here we
consider chirality due to different attachments of non-identical atoms
to the $C_\alpha$ atom.

\begin{figure}[hbt]
\centerline{\epsfig{figure=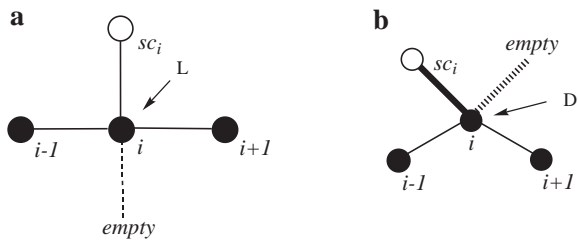,width=3in}}
\caption{\sf
Assignment of chirality for models
in planar square lattice and three dimensional tetrahedral
lattice. (a) A two-dimensional $L$-residue in square lattice.  From
the backbone monomer (filled circle) of residue $i+1$ to the side
chain atom ($sc_i$, empty circle) of residue $i$, then to the empty
site, we turn counter clockwise.  (b). A three-dimensional $D$ residue
in tetrahedral lattice.  Taking the view point along the vector
pointing from the backbone monomer of residue $i$ to the empty site,
the backbone atoms of residues $i+1, i-1$, and the side chain atom
($sc_i$) of residue $i$ are arranged in a clockwise fashion.
}
\label{Fig:ChiralModel}
\end{figure}

We first introduce chirality for two-dimensional lattice model.
Planar chirality arises if a two-dimensional molecule and its
reflection (mirror image about a line) cannot be superimposed.  For a
chiral residue, the placement of side chain atoms is restricted. The
chirality of a residue $i$ is determined by the relative orientation
of its attached side chain atom and the preceding and succeeding main
chain monomers of residues $i-1$ and $i+1$
(Figure~\ref{Fig:ChiralModel}a).  For a chiral residue $i$, if we
start from the main chain monomer of the succeeding residue $i+1$ and
go through the side chain atom of residue $i$ ($sc_i$) to the
unoccupied site (unoccupied by the two backbone monomers of residues
$i-1, i+1,$ and side chain $sc_i$), the chirality of residue $i$ is
$L$ if we turn counter clockwise, and $D$ if we turn clockwise.

For three dimensional tetrahedral lattice, chirality of a residue can
be defined realistically following that of the $C_\alpha$ in amino
acids (Figure~\ref{Fig:ChiralModel}b).  We take a view point along the
vector pointing from the backbone monomer of $i$ to the empty site
unoccupied by backbone monomers of residues $i-1$, $i+1$, and side
chain atom of residue $i$.  If the backbone monomers $i+1, i-1$, and
the side chain of residue $i$ are arranged counter clockwise, the
chirality of residue $i$ is $L$.  If they are arranged clockwise, the
chirality is $D$.  For achiral models in both square lattice and
tetrahedral lattice, there is no restriction for the allowed positions
for the first side chain atom and it can take any of the two available
sites unoccupied by backbone monomers of $i-1, i$, and $i+1$-th
residues.

\label{page:DL}
We study both chiral molecules and achiral molecules in this work.  In
a chiral molecule, the first atom of all side chains follow strictly
one fixed chirality (either D or L, we use D for this study).  In an achiral
molecule, there is no restriction and the first atom of a side chain
can take any unoccupied reachable site.

\paragraph{Models for side chain flexibility.}
\begin{figure}[hbt]
\centerline{\epsfig{figure=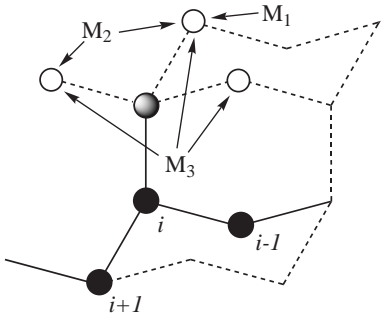,width=2.in}}
\caption{
\sf
Rotameric positions of $M_1$, $M_2$,
and $M_3$ side chain flexibility model.  Facing the vector pointing
from backbone atom $i$ to its first side chain atom, $M_1$ position
for the second side chain atom is located in the opposite direction to
the vector connecting backbone atom $i$ to backbone atom $i+1$.  $M_2$
model contains an additional site for the second side chain atom,
which is located in the opposite direction to the vector connecting
backbone atom $i$ to backbone atom $i-1$.  For $M_3$ model, the second
side chain atom can occupy any of the three reachable sites.
}
\label{Fig:FlexSC}
\end{figure}

Regardless whether a residue is chiral or achiral, it is possible to
have a flexible side chain if the side chain consists of two or more
atoms. Because square lattice and tetrahedral lattice both have
coordination number of four, there are at most three possible sites
available when attaching a new side chain atom.  This models well the
$\chi_1$ angels of protein side chains, as they often can be grouped
into mainly three clusters: $t$, $g^+$, and $g^-$, which stand for
trans, gauche positive , and gauche negative, respectively
\cite{DunbrackCohen97_PS}.  Depending on whether any of these sites
are forbidden, a side chain atom in tetrahedral lattice may have 1, 2,
or 3 allowed positions (Figure~\ref{Fig:FlexSC}). 
\label{page:rotamer}
In model $M_1$, the second
side chain atom can only be placed at one fixed position. In model
$M_2$, the second side chain atom can be placed at one additional
possible position, and in model $M_3$, the second side chain atom can
be placed at any of the three possible positions.  Facing the vector
pointing from backbone atom $i$ to its first side chain atom, $M_1$
position for the second side chain atom is located in the opposite
direction to the vector connecting backbone atom $i$ to backbone atom
$i+1$.  $M_2$ model contains an additional site for the second side
chain atom, which is the immediate neighbor of $M_1$ in
counterclockwise direction.  For $M_3$ model, the second side chain
atom can occupy any of the three reachable sites.

\paragraph{Contact and Compactness.}
\label{page:contact}
We focus on protein-like compact conformations.  Two non-bonded
monomer (backbone or side chain) $n_i$ and $n_j$ are in topological
contact if they are spatial neighbors.  Two main chain monomers are in
contact only if they are not sequential neighbor ( $i \ne j \pm 1$,
see arrows in Figure~\ref{Fig:LatModel}).

The parameter measuring
compactness $\rho$ of a conformation is defined as the ratio between
its number of topological contacts and the maximum number of contact
attainable for a particular sequence of given chain length
\cite{LauDill89_M}:
\[
\rho \equiv \frac{t}{t_{max}}, \quad \mbox{ where $0\le\rho\le1$.}   
\]

\paragraph{Entropy and excess entropy.}
We are interested in the effects of different models of side chain
chirality and flexibility on the conformational space of chain
polymers.  We study only homopolymers and do not investigate the
relationship between sequence and conformation.  

Since homopolymers do not fold into a unique stable ground state
conformation, we calculate the entropy for homopolymers to adopt
conformations at a specific compactness value $\rho$.  We
define entropy $S(\rho)$ for conformations with compactness $\rho$ as:
\[	 
	S(\rho)  = k_B \ln{n(\rho)} ,
\]
where $k_B$ is the Boltzmann constant, $n(\rho)$ is the number of
conformations with compactness $\rho$.  Similarly, side chain entropy
$S_{sc}(B)$ is defined for a fixed backbone conformation
$B$ as:
\[
	S_{sc}(B) = k_B \ln{n_{sc}(B)},
\] 
where $n_{sc}(B)$ is the number of all self-avoiding side chain
arrangements for the fixed backbone conformation $B$.
The overall entropy $S$ for all conformations is given by:
\[
	S= k_B \ln{\sum_i{n(\rho_i)}} = k_B \ln{\sum_j{n_{sc}(B_j)}}.
\]

The change $\Delta{S}$ 
in the conformational entropy 
between folded
state (F) and unfolded state (U) is given by:
\[
\Delta S = S_F - S_U.
\]
For lattice models used in this study, folded state is defined as
conformations with compactness $\rho = \rho_{max} \equiv 1$.  Unfolded
states correspond to all conformations with compactness $\rho < 1$.
We have:
\[
\Delta S(\rho_{max})= S(\rho_{max})-S(\rho<1).
\]
Since conformations with $\rho_{max}$ constitute a very small
proportion among all conformations: $S(\rho<1) \approx S$, 
we have:
\[
\Delta S(\rho_{max}) \approx S(\rho_{max})-S = k_B
\ln \frac{n(\rho_{max})}{\sum_i{n(\rho_i)}} = k_B\ln \omega(\rho_{max}),
\]
where $\omega(\rho_{max})$ is the fraction of maximum compact
conformations. For convenience, we define folding entropy
$\Delta{S_f}$ of the maximum compact conformations as the absolute
value of the above entropy change:
\[
	\Delta{S_f} =  |\Delta S(\rho_{max})| = -k_B\ln \omega(\rho_{max}).
\]
We define entropic change $\Delta S(\rho)$ at other compactness as:
\[
	\Delta{S(\rho)} =  |\Delta S(\rho)| = -k_B\ln \omega(\rho).
\]

To compare folding entropies of models with different chirality and
flexibility, we follow reference \cite{BrombergDill94_PS} and define
the excess entropy $ES_{a,b}$ for model $a$ when compared to model $b$
as:
\[	
	ES_{a,b} =
	\Delta S_f(a) - \Delta S_f(b) = -k_B \ln \frac{\omega_a(\rho_{max})}{\omega_b(\rho_{max})},
\]
where $\omega_a(\rho_{max})$ and $\omega_b(\rho_{max})$ are the
fractions of maximum compact conformations for model $a$ and model
$b$, respectively.

\paragraph{Radius of gyration $R_g$.} 

Radius of gyration ($R_g$) is a parameter frequently used to measure
the global compactness of a conformation.  For a set of $n$ atoms,
$R_g$ is the root-mean-square distance of position $\bx_i \in
\real^3$ of each atom $i$ to their geometric center $\bar{\bx} =
\sum_{i=1}^n \bx_i /n$:
\[
R_g= (\sum_{i=1}^n{(\bx_i- {\bar{\bx}})^2}/n)^{1/2}.
\]
For globular proteins, the value of $R_g$ fluctuates but can be
predicted with reasonable accuracy from the number of residues by the
relationship $R_g \approx 2.2N^{1/3}$, which describes accurately
globally compact proteins \cite{DimaThirumalai03_arXiv}.

\paragraph{Sequential Monte Carlo importance sampling.}
In this study, we need to estimate properties of rare events, namely,
properties of conformations with maximum number of contacts
$\rho_{max}$, {\it e.g.}, the fraction of conformations with
$\rho=\rho_{max}$.  Estimating properties of rare events is difficult,
because finding such conformations is challenging when more extended
conformations dominate in the whole population of all geometrically
feasible self-avoiding walks with side chains.  We adopt the same
sequential Monte Carlo strategy for sampling as that of a recent three
dimensional off-lattice study, where thousands of polymers of length
2,000 at very high compactness values were successfully generated
\cite{ZhangCTL03_JCP}.  Sequential Monte Carlo is an effective
strategy based on chain growth for sampling high dimensional space
\cite{Liu&Chen98,Liu01_MC}.  The details of studying lattice models
using this technique have been described elsewhere
\cite{LZC02_JCP,ZhangCTL03_JCP}.  It was shown previously that
sequential Monte Carlo can give accurate estimation of ensemble
properties of lattice conformations, as verified by comparison with
results obtained from exhaustive enumeration
\cite{LZC02_JCP,ZhangCTL03_JCP}.

Once a sample conformation is generated, we need to find out
whether it is maximally compact.  For two dimensional square lattice
models, the upper bound of the number of contacts $t^*$ for
polymers in which all beads (including main chain monomers and side
chain atoms) are connected can be calculated.  For any polymers with
$N$ beads, $t^*$ is:
\begin{equation}
\begin{split}
t^* &= N -2m,   \mbox{ for }   m^2 < N \leq m(m+1)\\
t^* &= N-1 -2m, \mbox{ for }   m(m+1) < N \leq (m+1)^2,
\end{split}
\end{equation}
where $m$ is a positive integer \cite{ChanDill89_MM}.  It is easy to
verify that this bound is tight for polymers without side chains and
gives the maximum number of contact $t_{max}$.

\begin{figure}[tb]
\centerline{\epsfig{figure=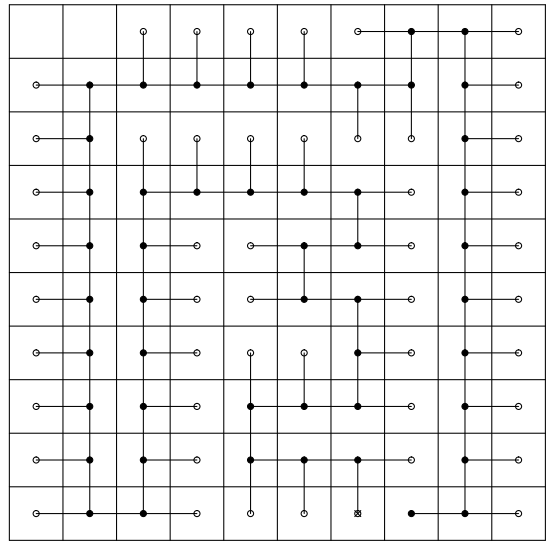,width=3in}}
\caption{\sf 
A maximum compact conformation on
square lattice for achiral model of side chain size one. Solid circles
are backbone monomers, and empty circles are side chain atoms.  The
main chain length is 50.
}
\label{Fig:MaxExmp} 
\end{figure}

Finding the maximum number of contacts for models with side chain is
more difficult, since no closed-form answers are known for various
side chain models studied here. The compactness $\rho$ is therefore
difficult to calculate for long chain polymers.  With the introduction
of side chains, it is possible that the maximum compact conformations
may not take $t^*$ as $t_{max}$ due to the requirement of side
chain connectivity and self avoidance.  For two dimensional square
lattice models with side chain of size one, we find from exhaustive
enumeration that there are conformations with maximum contact of
$t^*$ for chains up to length $N=18$ in achiral model. No
conformations with $t^*$ contacts exist for chiral model.  For longer
chains, we generate samples of conformations using sequential Monte
Carlo for length up to $N=100$.  We found that for achiral models,
there exist sampled conformations with $t^*$ contacts at every length from
19 to 100.  This suggests that it is likely that achiral models with
side chain of side one have $t^* = t_{max}$ at length $N \geq
2$. It also indicates that this sampling strategy is effective and our
method can give correct estimation of the maximum number of contact
$t_{max}$ and compactness $\rho$ for two dimensional achiral model.
An example of maximum compact conformation for achiral model of length
50 on square lattice obtained using sequential Monte Carlo is shown in
Figure~\ref{Fig:MaxExmp}, with $N= 98$ and $m= 9$.

Verified successful results in two-dimensional models are helpful in
assessing the effectiveness of sampling for three-dimensional models.
Both tetrahedral lattice models and square lattice models have the
same coordination number of four. In addition, conformations from
chiral model is a subset of that of achiral model. We postulate that
our method can give satisfactory estimation of $\rho_{max}$ for
tetrahedral models used in this study.

\paragraph{Exact calculation of side chain entropy.}
Rotamer counting is a widely used method to estimate side chain
entropy of residues in proteins when the backbone structure is given
\cite{PickettSternberg93_JMB}.  The idea is to count available
rotameric states for each monomer independently, and estimate the
total number of states by multiplication. This approach would be
accurate if all possible placement of side chains at different
residues are independent. The problem is that not all combinations of
rotameric states for residues along the main chain are self-avoiding.
Hence this method inherently overestimate conformational entropy.  The
extent of the over-estimation and its effect in assessing protein
folding entropy is unknown.

Calculating the exact number of all valid side chain conformations for
a given main chain structure is challenging, since this requires
explicit enumeration of all possible spatial arrangement of side
chains.  Here we introduce an algorithm for counting side chain
conformations based on the divide-and-conquer paradigm.  

\begin{figure}[hbt]
\centerline{\epsfig{figure=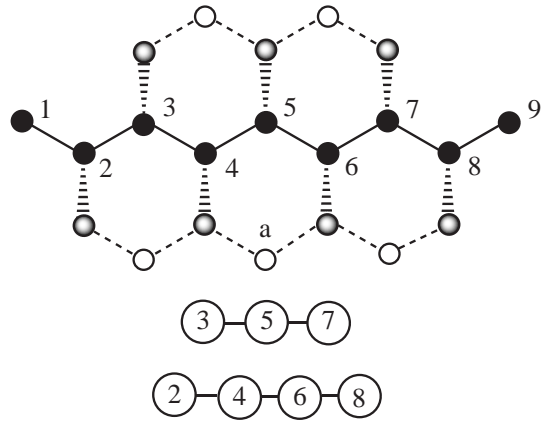,width=2.8in}}
\caption{\sf
An illustration of calculation of side
chain entropy.  (a) An extended conformation.  Filled circles are main
chain monomers, gray circles are first side chain atoms, and empty
circles are positions that could be occupied by side chain atoms of
either of two different residues. (b) The conflict graph of the
conformation.  There are two disconnected components in the conflict
graph, one formed by residues 3, 5, and 7, and another by residues 2,
4, 6, and 8.  The latter can be further divided into two smaller
components by cutting at position $a$.
}
\label{Fig:flx_alg}
\end{figure}

For a fixed backbone structure, if the placement of side chain atoms
of a residue affects the allowed positions of side-chain atoms of
another residue, we say there is a conflict for side chains of these
two residues.  We can construct a conflict graph $G = (V, E)$, where
$V$ is the set of residues, and $E$ is the set of edges representing
conflicts between pairs of residues.  All residues in a molecule can
be grouped into $m$ individual sets, each representing a disconnected
component of the conflict graph $G$.  When two sets are disconnected,
side-chain placement of residues in one set does not affect the
placement of side-chains of residues in another set.  The disconnected
components in graph $G$ can be identified using depth-first-search
\cite{Cormen90}.  We can then calculate the number of different
side-chain arrangement $n_i$ for each set $i$ by enumeration.  The
total number $N$ of side chain conformations of all residues is
obtained by multiplication: $N = \prod_{i=1}^m n_i$.  A simple example
is shown in Figure~\ref{Fig:flx_alg}a and Figure~\ref{Fig:flx_alg}b, where a
graph is constructed for an extended backbone structure.  The residues
can be decomposed into two independent components, one formed by
residues with side chains above the main chain, and another formed by
residues with side chains below the main chain.

\paragraph{Helix content on tetrahedral lattice.}

\begin{figure}[tb]
\centerline{\epsfig{figure=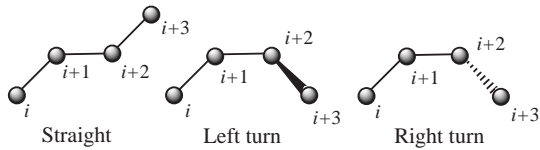,width=2.8in}}
\caption{\sf The three possible conformations of
a four-fragment backbone conformations.
}
\label{fig:Helix-exp}
\end{figure}

For a fragment of four consecutive monomers (from $i$ to $i+3$), there
are three possible conformations on a tetrahedral lattice: left turn
fragment, right turn fragment, and straight fragment (see
Figure~\ref{fig:Helix-exp}).  For a fragment of five consecutive
monomers, the 4-prefix fragment ($i$ to $i+3$) and the 4-suffix
fragment ($i+1$ to $i+4$) can have any of the above three
conformations.  If the 4-prefix and 4-suffix fragments are of all left
turns or all right turns, this five monomer fragment is defined as a
helix.  Helices with all left turns are defined as left-hand helices,
and helices with all right turns are defined as right-hand helix. We
include both types of helices and their mixture when calculating helix
content of a backbone conformation.  Specifically, the helix content
$h$ of a backbone conformation of length $N$ is:
\[
h = \sum_{i=0}^{N-5} \Bbb{I}(i)/(N-4),
\]
where $\Bbb{I}(i) =1$ if the fragment of residues from $i$ to $i+4$ is
a helix by the above definition, and $\Bbb{I}(i) =0$ otherwise.

\section{Results}

\subsection{Exact conformational space by enumeration}

\begin{table*}[thb] % one column table
 \begin{center}
\caption{\sf
Number of conformations of a
$n$-polymer by enumeration for different side-chain models.  Here
``2Z'' stands for two-dimensional square lattice, ``3T'' for
three-dimensional cubic lattice, ``c'' for chiral models, ``a'' for
achiral models, ``1'' and ``2'' for side-chain models of 1 and 2
atoms, respectively, ``M1--M3'' for specific models of side-chain
flexibility, where the second side chain atom can have 1--3 allowed
positions, respectively.  Specifically, we have: 2Za1: square lattice
achiral model with side chain size of one, 2Zc1: square lattice chiral
model with side chain size of one, 3Ta1: tetrahedral lattice achiral
model with side chain size of one, 3Tc1: tetrahedral lattice chiral
model with side chain size of one, 3Tc2.M1--M3: tetrahedral lattice
chiral model with side chain size of two and flexibility models of
M1--M3, respectively.
}
\label{tab:enumeration}
\vspace*{.1in}
 \begin{tabular}{rrrrrrrr}
\hline 
   $n$ & 2Za1 & 2Zc1 & 3Ta1 & 3Tc1 & 3Tc2.M1 & 3Tc2.M2 & 3Tc2.M3   \\ \hline
3  &    6      &         3  &     2     &          1    &           1     &          2      &         3  	\\
4   &    36      &        9   &    12      &        3      &         3         &      12           &   27	\\
5    &   152     &        19  &     72      &         9     &           9   &             70  &             237	\\
6  &     688      &       43    &   432       &       27      &         25    &           377   &           1888	\\
7   &    2784     &       86    &   2336        &     73        &       63      &         1820    &         13659	\\
8    &   11744      &      182   &   12992        &    203        &      158      &        8784     &        98202	\\
9  &     47488        &    360   &   70720          &  553          &    386        &      41002    &       692820	\\
10  &    195872         &  740    &  388096        &   1519           &  931  &            189167    &      4833081	\\
11  &    791552      &     1453   &  2095872      &    4109 &            2220   &          859214     &     33447567	\\
12   &   3233568       &   2930   &  11392416      &   11179  &          5309     &        3913808     &    231456640	\\
13   &   13046720     &    5698   &  61468544    &     30240    &        12695   &         17752390     &   1596526404	\\
14   &   53015776      &   11343  &  332851456     &   82021      &      30281  &          80385077  &      11002320270	\\
15   &   213565776   &     21847  &  1792133312      & 221401       &    72159   &         363404876  &     75735208118	\\
16   &   864828096    &    43072  &  9674958976   &    598996         &  171914    &       1642367812  &       NA	\\
17   &   3478827632    &   82297  &  52031751936    &  1614693  &        409056      &     7413318612    &      NA	\\
18   &   14051949392   &   160938  & 280368151936  &   4360282    &      972797        &   33449593868     &    NA	\\
19   &   NA           &    305280    &  NA         &   11741404     &    2311751   &          NA    &           NA	\\
20   &   NA           &    592686     & NA          &  31661162       &  5491818     &        NA      &         NA	\\
\hline 
 \end{tabular}
 \end{center}
\end{table*}

For short polymers with side chains, we obtain a complete picture of
the ensemble properties of conformations by exhaustive enumeration.
Table~\ref{tab:enumeration} lists the total number of conformations of
different side chain models obtained from exhaustive enumeration.  For
longer polymers, the full conformational space cannot be enumerated,
and it is necessary to use sequential Monte Carlo sampling to generate
properly weighted samples from the uniform distribution of all
geometrically feasible self-avoiding walks with various types of side
chains.

\subsection{Effects of side chain chirality}
\paragraph{Distribution of conformations and folding entropy.}
How does the introduction of chirality affect the distribution of
conformations and the entropy of folding?  We first calculate the
distributions of conformations over compactness $\rho$ for
a given main chain length $N$. The fraction $f(\rho)$ of conformations
with compactness $\rho$ is:
\[
f(\rho) \equiv \frac{\omega(\rho)}{\sum_\rho\omega(\rho)},
\] 
where $\omega(\rho)$ is the number of conformations found with
compactness $\rho$. 

\begin{figure}[tb]
\centerline{\epsfig{figure=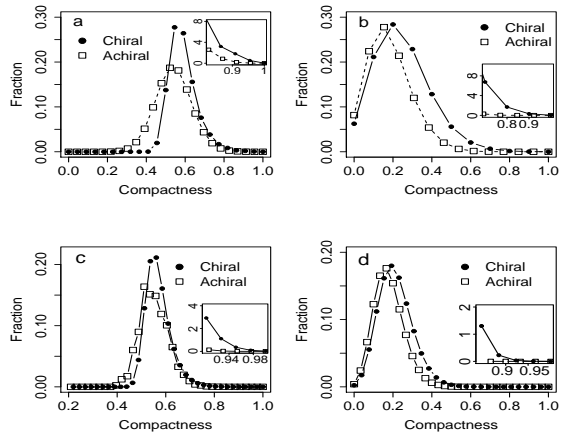,width=3in}}
\caption{\sf Distribution of conformations over
compactness for models of side chain size one (a) polymer chains of
length 16 on two-dimensional square lattice (2Za1 and 2Zc1) obtained
by exhaustive enumeration, (b) polymer chains of length 16 on
three-dimensional tetrahedral lattice (3Ta1 and 3Tc1) obtained by
exhaustive enumeration, (c) polymer chains of length 30 on
two-dimensional square lattice obtained by sequential Monte Carlo, and
(d) polymer chains of length 30 on three-dimensional tetrahedral
lattice obtained by sequential Monte Carlo.
}
\label{Fig:ch_dist}
\end{figure}

The distributions of enumerated conformations of chiral and achiral
polymers at length 16 on two-dimensional square lattice and
three-dimensional tetrahedral lattice are shown in
Figures~\ref{Fig:ch_dist}a and ~\ref{Fig:ch_dist}b, respectively.  The
distribution of conformations at length 30 estimated by sequential
Monte Carlo for both square and tetrahedral lattice are shown in
Figures~\ref{Fig:ch_dist}c and ~\ref{Fig:ch_dist}d, respectively.  
\label{page:insert}
The inserts provide details of conformations in compact region.
Distributions of less compact conformations may be useful for modeling
proteins in unfolded state.  Results from enumeration and sampling
show similar patterns. In both two and three dimensional space, chiral
and achiral polymers have low average compactness when the only
interaction between residues is due to excluded volume, as in good
solvent. However, the distributions of conformations of these two
chirality models are clearly different.  Chiral models have overall
more compact conformations than achiral models.

\begin{figure}[thb]
\centerline{\epsfig{figure=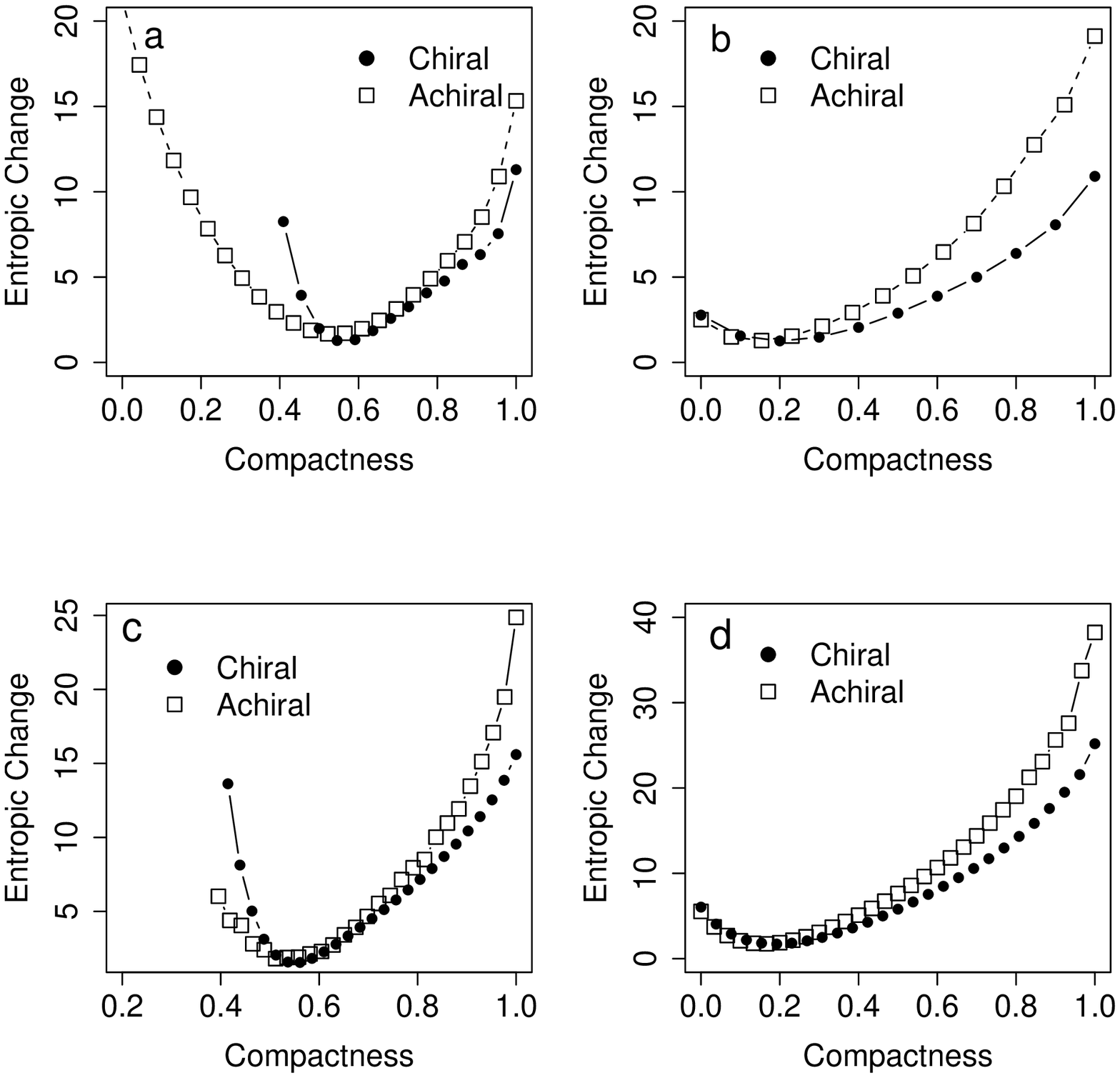,width=3in}}
\caption{\sf
Folding entropy for different models of
chirality with side chain size of one.  (a) Exact folding entropy
$\Delta S_f$ at maximum compactness and entropic change $\Delta
S(\rho)$ at other compactness regions $\rho$ for polymers of length 16
calculated by exhaustive enumeration on two-dimensional square
lattice, (b) Exact folding entropy $\Delta S_f$ and entropic change
$\Delta S(\rho)$ for polymers of length 16 calculated by exhaustive
enumeration on three-dimensional tetrahedral lattice, (c) Estimated
$\Delta S_f$ and $\Delta S(\rho)$ for polymers of length 30 calculated
by sequential Monte Carlo on two-dimensional square lattice.  and (d)
on three-dimensional tetrahedral lattice.
}
\label{Fig:ch_fe}
\end{figure}

Since proteins are highly compact, we consider conformations in high
compactness region, especially in the region where $\rho = 1$. We
calculate the folding entropy $\Delta S_f$ for the ensemble of
conformations with maximum compactness $\rho=1$ and entropic change
$\Delta S(\rho)$ at other compactness region.
Figures~\ref{Fig:ch_fe}a and ~\ref{Fig:ch_fe}b show exact $\Delta S_f$ and
$\Delta S(\rho)$ for two and three dimensional polymers of length 16
calculated by enumeration.

For chiral models, the change in entropy during folding to
conformations of maximum compactness is much smaller than that of
achiral models.  For chiral and achiral conformations on tetrahedral
lattice with side chain size 1 (3Tc1 and 3Ta1 in
Table~\ref{tab:enumeration}), the fraction of maximum compact
conformations is much higher for chiral molecules ($1.8 \times
10^{-5}$) than for achiral molecules ($4.96\times 10^{-9}$) at length
16.  Chirality clearly favors compact conformations, despite the fact
that the absolute number of conformations of maximum compactness is
much smaller for chiral model (11 conformations) than for achiral
models (144 conformations).

\paragraph{Excess folding entropy  due to chirality.}

\begin{figure}[tb]
\centerline{\epsfig{figure=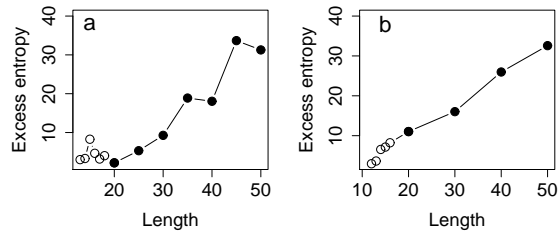,width=3.in}}
\caption{\sf
Excess folding entropy due to chirality
for maximum compact conformations calculated for models of side chain
size one, (a) on two-dimensional square lattice, and (b) on
three-dimensional tetrahedral lattice.  Unfilled circles are excess
folding entropy calculated by exact enumeration. Filled circles are
excess folding entropy estimated by sequential Monte Carlo.
 }
\label{Fig:ch_ee}
\end{figure}

To examine the differences of folding entropy for polymers under two
different chirality models, we calculate excess entropy of folding
$ES_{c, a}(\rho)$ of chiral model over achiral model for conformations
at maximum compactness $\rho = 1.0$.  The scaling relationships of
$ES_{c,a}{(1.0)}$ with polymer backbone length $N$ are shown in
Figures~\ref{Fig:ch_ee}a and ~\ref{Fig:ch_ee}b for square lattice model
and tetrahedral lattice model, respectively.  It
provides information on whether and how the effects of chirality
changes with chain length.

Exact $ES_{c,a}(1.0)$ obtained by enumeration in square lattice up to
$N=18$ fluctuates with chain length (Figure~\ref{Fig:ch_ee}a, unfilled
circles).  In tetrahedral lattice, $ES_{c,a}$ increases with $N$ to
8.2 at length $N=16$ (Figure~\ref{Fig:ch_ee}b, unfilled circles).
This trend becomes clearer in results obtained by sequential Monte
Carlo for conformations up to length $N=50$ (Figure~\ref{Fig:ch_ee}b).
When chain length increases, the excess entropy increases
linearly. This suggests that the effect of chirality on entropy of
folding increases with chain length in tetrahedral lattice. For
tetrahedral lattice, this relationship can be characterized by a
linear regression ($R^2 = 0.98$) with $ES_{c,a}(1.0, N) = a N + b$,
with $a = 0.75 \pm 0.06$, $b = -4.7 \pm 2.4$. The effects of chirality
in increasing the fraction of compact chains become more pronounced as
chain length increases.  In square lattice the effect of chirality to
excess folding entropy also increases with chain length, but the trend
is not as clear as that in tetrahedral lattice.

\subsection{Effects of side chain flexibility}
Because natural amino acid residues are three dimensional chiral
molecules, we describe only results on flexibility effects using 
chiral models on three-dimensional tetrahedral lattice.  A benefit
from studying chiral model is that the conformational space of side
chain is greatly reduced, and the folding entropy can be studied for
longer polymers.  We omit results on two dimensional square lattice,
which are similar to that of tetrahedral lattice shown here.

\paragraph{Distribution of conformations and folding entropy.}

\begin{figure}[tb]
\centerline{\epsfig{figure=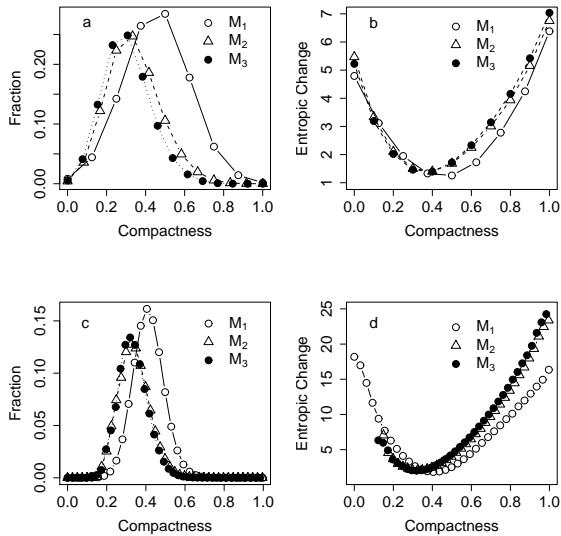,width=3in}}
\label{Fig:flx_smc}
\caption{\sf
Distribution of conformations and
folding entropy over compactness for models with different side chain
flexibility.  Exact distribution of (a) conformations and (b) folding
entropy for chains of length 12 obtained by exhaustive enumeration,
and estimated distribution of conformations (c) and folding entropy
(d) for chains of length 30 obtained by sampling.
  }
\label{Fig:flx_dist_fe}
\end{figure}

To study the effect of side chain flexibility, we first examine the
exact distributions of conformations obtained by enumeration for
polymers of length $N=12$. These are obtained for three different
models $M_1, M_2$, and $M_3$ of different side chain flexibility,
where the second side-chain atom can have 1, 2, and 3 allowed
positions, respectively.  The distributions of conformations of the
three models show that $M_1$ has much higher average compactness
compared to the other two models (Figure~\ref{Fig:flx_dist_fe}a). That
is, less flexible side chains are more likely to form compact
conformations. $M_1$ model also has the lowest folding entropy for
compact conformations (Figure~\ref{Fig:flx_dist_fe}b). Model $M_2$ and
$M_3$ have similar distribution, with $M_2$ slightly more compact on
average.

For polymers of chain length $N=30$, the distributions of
conformations estimated by sequential Monte Carlo show a similar
pattern. Conformations from model $M_1$ on average
are more compact (Figure~\ref{Fig:flx_dist_fe}c): the largest number
of conformations are found around $\rho \approx 0.42$ compared to
$\rho \approx 0.34$ for model $M_2$ and model $M_3$.  For compact
conformations ({\it e.g.}, $\rho \ge 0.8$), the entropic change is
also much smaller for $M_1$ model of inflexible side chains.  This
suggests that there is a significant decrease in folding entropy when
side chain loses its flexibilities.  

\paragraph{Excess folding entropy due to side chain flexibility.}
\begin{figure}[tb]
\centerline{\epsfig{figure=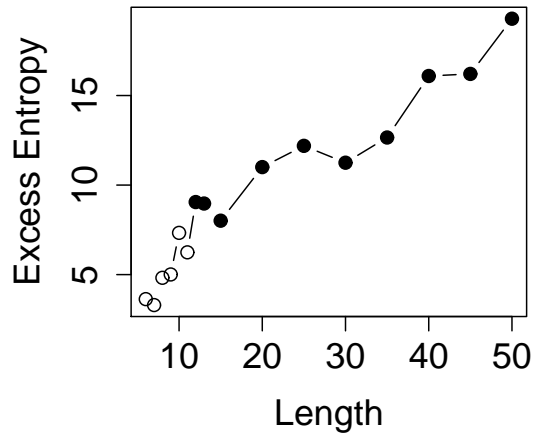,width=3in}} 
\caption{\sf Excess entropy of folding for
conformations from $M_3$ model over conformations from $M_1$ model
estimated by sequential Monte Carlo.
 }
\label{Fig:flx_ee} 
\end{figure}

The excess entropy of folding $E_{M_1, M_3}(1.0)$ of model $M_1$
compared to model $M_3$ for maximum compact conformations is shown in
Fig~\ref{Fig:flx_ee}. It can be characterized by a linear regression
$ES_{M_1,M_3}(1.0, N) = a N + b$, with $a = 0.27 \pm 0.03$, $b = 4.75
\pm 1.0$, and $R^2 = 0.90$.  These results suggest that inflexibility
of side chain plays an important role for obtaining compact
conformations.  The effects of inflexibility in increasing the
fraction of compact chains become more pronounced as chain length
increases.

\subsection{Effects of side chains: packing, entropy, and secondary structure.}
\paragraph{Jigsaw puzzle or nuts and bolts?}
\begin{figure}[tb]
\centerline{\epsfig{figure=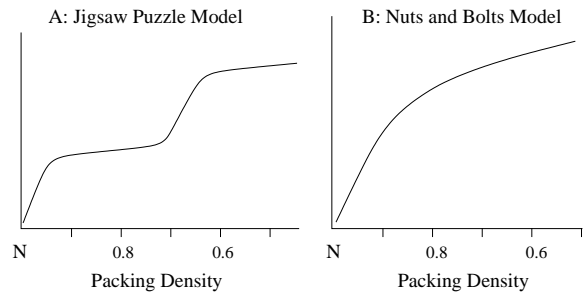,width=3in}} 
\caption{\sf A schematic comparison shows the
qualitative difference in the dependencies of side-chain entropy on
chain density in (a) a jigsaw puzzle model for side-chain packing, in
which a side chain freezing effect occurs at near compact region; and
(b) a nuts and bolts model in which main chain and side chain degrees
of freedom are linked. (Adapted from Fig 13 in
\cite{BrombergDill94_PS})
}
\label{Fig:JB_exp} 
\end{figure}

Two differing views on the effects of side chains can be summarized by
the model of jigsaw puzzle and the model of nuts and bolts
(Fig~\ref{Fig:JB_exp}).  This
comparison was studied in details in the seminal work of reference
\cite{BrombergDill94_PS}, where homopolymers of side chain of size one
are studied.  According to the nuts and bolts model, a small expansion
in the volume of compact native protein leads to a large increase in
side-chain entropy.  That is, side-chain entropy increases sharply as
main chain becomes less packed than native state.  According to the
jigsaw puzzle model, a small expansion in volume does not lead to
significant change in side chain entropy when the molecule is
compact. In a model supporting the jigsaw puzzle mode, it is estimated
that a 25\% expansion in volume relative to the native core volume is
required before a sudden unfreezing of core side-chain rotameric
degrees of freedom incurs a sharp increase in entropy
\cite{ShakhnovichFinkelstein89_Biop}. In a model supporting the nuts
and bolts model, a small expansion in volume from the compact native
state produces a steep increase in side-chain rotational entropy
\cite{BrombergDill94_PS}.  The increase in side-chain degrees of
freedom is linked to the increase in main-chain degrees of freedom
\cite{BrombergDill94_PS}.  In this study, the size of the side chain
is 1.  For real proteins, except glycine and alanine, all other amino
acid residues have more than one heavy atom in their side chains.
What effects do side chains of larger sizes have on side chain
packing?  With the method of exact computation of side chain entropy,
we revisit this problem and examine the packing of chiral polymers
with side chains formed by two atoms ($M_3$ model).  Following
reference \cite{BrombergDill94_PS}, we use the radius of gyration
$R_g$ of backbone monomers to measure main chain packing density.

\begin{figure}[tb]
\centerline{\epsfig{figure=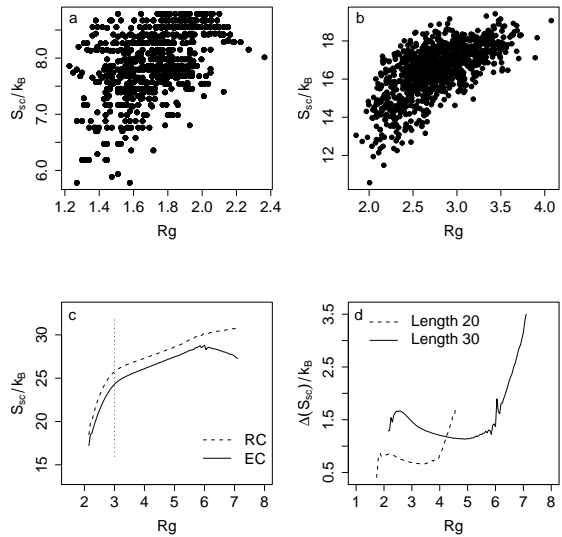,width=3in}} 
\caption{\sf
Side chain entropy and main chain
compactness of tetrahedral chiral conformations with two-atom side
chains, where the second atom can have three allowed positions.  (a)
Exact number of side chain conformations for all main chain structures
of length 10 over the main chain radius of gyration ($R_g$), (b) Exact
number of side chain conformations for a set of 1,000 samples of main
chain structures of length 20 over $R_g$, c) Expected side chain
entropy of main chain structures of length 30 at different radius
gyration, as estimated from exact calculation of side chain entropy
based on 1,000,000 properly weighted samples of main chain structures
(EC, solid line). Resampling technique described in \cite{LZC02_JCP}
was used to obtained samples in each intervals of $R_g$ values. For
comparison, side chain entropies estimated by rotamer counting are
also plotted (RC, dotted line), and (d) Difference in expected side
chain entropy by exact calculation and rotamer counting from sampled
backbone structures at length 20 (solid line) and 30 (dotted line).
 }
\label{Fig:scEn_rg} 
\end{figure}

We examine the distribution of side chain entropy $S_{sc}$
over the full range of main chain packing density measured by $R_g$.  Exact
calculation of side chain entropy for each of the exhaustively
enumerated main chain conformation of length 10 shows that side chain
entropy does not always correlate well with main chain packing density
(Figure~\ref{Fig:scEn_rg}a).  Conformations with most extended main
chain structures ($R_g = 2.2 - 2.4$) are not those with maximum side
chain entropy, and many compact conformations ($R_g \approx 1.4$) have very
large side chain entropy.

We then use sequential Monte Carlo to generate longer main chain
structures up to length 30, and calculate exactly the side chain
entropy for each of the sampled main chain structures.  We assess the
correlation of side chain entropy and main chain backbone packing by
calculating the average side chain entropy for conformations whose
backbone $R_g$ value falls into different intervals.  As shown in
Figure~\ref{Fig:scEn_rg}b, although some of the compact main chain
structures of length 30 have very small $Rg$, they can still have
substantial side chain entropy.

On average, there is a sharp decrease in the number of side chain
conformations at compact regions where main chain $R_g$ values are
small for polymers of chain length $N=20$.  There is no plateau at
compact regions with small $R_g$ value, which would be characteristic
of the jigsaw puzzle model.  Our study using chiral model of
homopolymers with two side chain atoms therefore is consistent with
the nuts and bolts model of protein packing.

\paragraph{Rotamer counting.}
Estimating side chain entropy is an important problem that received
much attention
\cite{Creamer00_P,BradySharp97_COSB,DoigSternberg95_PS,YuHodges99_JACS}.
For example, it was proposed in ref.\ \cite{YuHodges99_JACS} that
side-chain entropy should be used as a criterion alternative to
packing density to assess protein packing.  Models developed in this
study allow us to calculate explicitly side-chain entropy. We compare
the numbers of side chain conformations obtained by exact calculation
and by estimation using rotamer counting.  With sequential Monte
Carlo, we can access polymers in the full range of main chain
compactness, including both maximum compact backbones and fully
extended backbones, as well as polymers with compactness in-between.
Because each sampled conformation is properly weighted, we have thus
an accurate picture of the full distribution of all feasible geometric
conformations for various side chain models.  This is different from
other approaches such as molecular dynamics, where one typically
samples conformations around the native structure
\cite{SchaferSmithMarkGunsteren02_P}.

We find that the number of side chain conformation by rotamer counting
is consistently higher than the number obtained from exact enumeration
(Figure~\ref{Fig:scEn_rg}c and \ref{Fig:scEn_rg}d).  The difference
between these two methods varies at different main chain compactness.
Over-estimation by rotamer counting is especially large for very
extended conformations. It is also pronounced near the maximum compact
region. That is, there is substantial unaccounted effect of side chain
correlation in reducing side chain entropy due to excluded volume for
rotamer counting , and this effect is more pronounced in both extended
and near maximum compact regions.

\paragraph{Side chain entropy, compactness, and secondary structures.}
\begin{figure}[tb]
\centerline{\epsfig{figure=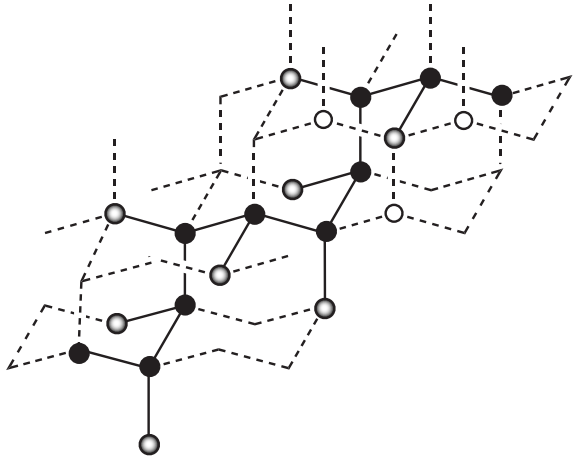,width=3in}} 
\caption{\sf
Side chain entropy and compactness
constratint favor the formation of helix-like structures.
An example of main chain structure of length 10 with
maximum side chain entropy is shown here. It adopts a helix-like
conformation.
}
\label{Fig:flx_exp} 
\end{figure}

How does side chain entropy affect the formation of secondary
structures?  An example of a conformation at length $N=10$ from
tetrahedral chiral $M_3$ model (two side chain atoms with three
possible position for the second atom) with maximum number of possible
side chain conformations is shown in Figure~\ref{Fig:flx_exp}.  Since
the second side chain atom in $M_3$ model can have three possible
sites, side chain entropies of different residues may be correlated
due to excluded volume effect if their side chain atoms can reach the
same lattice site.  The backbone structure of this particular
conformation is arranged in such a way that none of the second atoms
from different side chains can occupy the same lattice site.  That is,
in the conflict graph of this backbone, all vertices representing
individual residues are disconnected, and there are $N=8$ independent
components in the graph.  There is no correlation between side chain
entropies of any residues in this backbone structure, and the total
side chain entropy is simply determined by the total number of states
of side chain $\prod_{i=1}^N n_i$, where $n_i = 3$ for $M_3$ model.
It is remarkable that the spatial arrangement of this backbone
structure resembles that of a helix.  This suggests that the formation
of helical secondary structures is strongly favored by side chain
entropy for compact conforamtions.

\begin{figure}[tb]
\centerline{\epsfig{figure=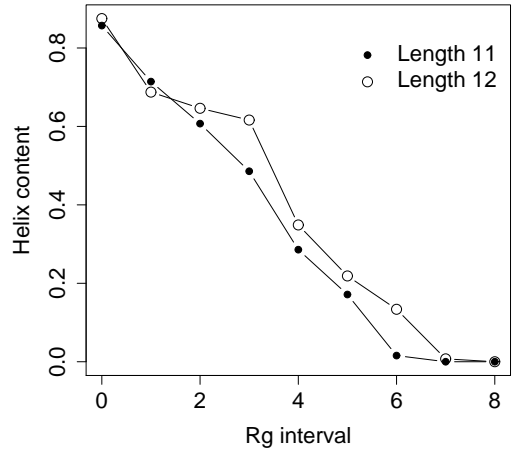,width=2.8in}}
\caption{\sf
Distributions of mean helix content
for backbone conformations of $M_3$ model with maximum side chain
entropy at different compactness for chains of length $N=11$ and
$N=12$ as measured by radius of gyration $R_g$.  We use 9 bins of
uniform width for backbone conformations whose $R_g$ values are all
within the range of 1.8 to 2.4 lattice unit length.  Compact backbone
conformations have higher helix content ({\it e.g.}, the 0-th and the
1-st bins).
}
\label{fig:helixHist}
\end{figure}

In contrast, the most extended backbone has much smaller side chain
entropy (Figure~\ref{Fig:flx_alg}).  Although the second atom of all
side chain can have three possible positions, an empty site reachable
from two residues can be taken by the side chain of only one residue,
and the total number of possible states for side chains is much
smaller.  The conflict graph of the backbone structure in
Figure~\ref{Fig:flx_alg} has only two independent disconnected
components, one formed by residues whose side chains are pointing up,
and another formed by those whose side chains are pointing down.  We
found that the mean helix content for backbones with maximum side
chain entropy increases rapidly as backbones become more compact.
Intra-chain hydrogen bond has long been thought as the determining
factor for the formation of helical secondary structures. Side chain
entropy was shown as an opposing factor for helix formation by
molecular dynamics simulation \cite{CreamerRose92_PNAS}.
Results obtained here show that the combination of side chain entropy
and compactness constraint lead to preference of helix formation
(Figure~\ref{fig:helixHist}).  
Helices have been observed
in polyproline molecules as small as three to five residues on the
basis of vibrational and ultraviolet CD measurements
\cite{DukorKeiderling91_BP}.  Polyproline molecules cannot form
intrachain hydrogen bond.  Experimental results on polypeptoids also
suggest that side chain entropy can lead to formation of significant
helical structures \cite{KirshenbaumZuckermann98_PNAS}.  For example,
artificially synthesized polypeptoids lack amide protons and are
incapable of forming intra-chain hydrogen bonds. However, they can
form monomeric alpha helices, as evidenced by CD spectra studies of
pentameric and octameric peptoids.  These alpha helices display
characteristics of peptide behavior such as cooperative pH- and
temperature-induced unfolding in aqueous solution.  Examination of the
structural details of these artificially synthesized polypeptoids
indicates that side chain steric interaction in extended conformations
of backbone is effectively avoided in helical conformations, as shown
in our model study.  Excluded volume effect of side chains leads to
preference of helical backbone conformations over extended backbone
conformations.  This consideration may be useful for rational design
of foldable polymer.

\section{Discussion}
In this study, we have developed two and three dimensional lattice
models with explicit side chains.  We make the distinction between the
effects due to side chain chirality and to side chain flexibility.  Side
chains do not readily convert between configurations of different
chiralities, whereas flexible side chains with two or more atoms can
easily take different rotameric states when spatially feasible.  We
examine specifically the effects of side chain chirality and side
chain flexibility on the distribution of polymers at different
compactness, and their effects on folding entropy.

We find that polymers from chiral models on average are more compact
than those from achiral models.  Chiral models also have significantly
smaller folding entropy into compact conformations than achiral
models. The excess folding entropy between achiral models and chiral
models increase linearly with chain length for long chains, suggesting
the effects of chirality becomes more important for long chain
polymers.

We also find that models with less flexible side chains have lower
entropy of folding than those with more flexible side chains.
Polymers with more flexible side chains may be thought to have better
chance to fit into a compact state.  However, there is a large
entropic cost associated with flexible side chains. The excess entropy
of flexible over inflexible side chain models also increases with
chain length. These findings suggest that amino acid residues in
proteins need to maintain a reduced flexibility to ensure fast folding
and stable native structure.

With explicit side chains, our study also confirms the conclusion of
an earlier study based on simpler side chain models, namely, side
chain packing is more like nuts-and-bolts rather than jigsaw puzzle,
and main chain and side chain degrees of freedom are linked.

\begin{table*}[thb] % one column table
 \begin{center}
\caption{\sf
Side chain flexibility of naturally
occurring amino acid residues.  H: hydrophobic residues, P:
hydrophilic or polar residues, N: neutral residues, $f = N_r/N_a$:
number of rotamers per atom. Values in parenthesis after residue names
are hydrophobicity values of amino acid residues as given in
\cite{BlackMould91_AB}, and values in parenthesis after $f=N_r/N_a$
values are the relative frequency of occurrence of the corresponding
amino acid residue in percentage \cite{TekaiaYeramianDujon02_G}.
}
\label{tab:Nr_Na}
\vspace*{.2in}
 \begin{tabular}{lr|lr|lr}
\hline 
H    & $N_r/N_a$ & N    & $N_r/N_a$ & P    & $N_r/N_a$  \\ \hline
F(1.0)  & 0.57(4.29)  & A(0.62) & 1.00(6.24) & D(0.028) & 1.25(5.38)  \\ 
I(0.94) & 1.75(5.80)  & C(0.68) & 1.50(1.70) & E(0.043) & 1.60(6.57)  \\
L(0.94) & 1.25(9.40)  & G(0.50) & --(5.64) & H(0.165) & 1.33(2.33)  \\
V(0.83) & 1.00(5.99)  & M(0.74) & 3.25(2.21) & K(0.283) & 5.40(6.46)  \\
W(0.87) & 0.70(1.11)  & P(0.71) & 1.00(4.96) & N(0.236) & 1.75(5.08)  \\
Y(0.88) & 0.50(3.15)  & S(0.36) & 1.50(8.67) & Q(0.251) & 1.80(4.23)  \\
        &	      & T(0.45) & 1.00(5.60) & R(0.000) & 4.86(4.99)  \\ \hline
Average	& 1.10(29.74) &         & 1.34(35.02) &          & 2.74(35.04) \\ \hline
 \end{tabular}
 \end{center}
\end{table*}

It is informative to examine the side chain flexibility of natural
amino acid residues. Among the 20 amino acids, all non-polar amino
acid residues either have branched side chains, or are aromatic with
ring structures. On average they are rather inflexible. The total
number of rotatable bonds divided by the number of side chain atoms is
small (Table~\ref{tab:Nr_Na}).  In contrast, side chains of polar or
ionizable residues such as lysine and arginine have more rotatable
bonds and have higher flexibilities.  However, this difference can be
rationalized by the observation that side chains of polar and
ionizable residues often are involved in electrostatic ion pair
interactions or hydrogen bonding interactions when buried in protein
interior, hence they have effectively reduced flexibility. The overall
flexibilities of side chains of all natural residues are therefore
relatively small. This reduced flexibility may be necessary to
decrease the entropy opposing folding to compact state.

Examination of patterns of side chain rotamer libraries further
confirms this observation.  We use a parameter $f$ for the number of
rotamers per atom defined as $f= n_r/n_a$, where $n_r$ is the number
of all possible rotamers for a specific side chain type and $n_a$ is
the number of heavy atoms in that side chain type. By the criterion of
hydrophobicity \cite{BlackMould91_AB}, 
we divide twenty amino acids into three categories: hydrophilic 
residues (hydrophobicity $<$ 0.3), hydrophobic residues 
(hydrophobicity $>$ 0.75), and neutral residues (0.3 $\geq$ hydrophobicity $\leq$ 0.75). 
According to this division, there are seven hydrophobic residue types, 
seven hydrophilic residue types, and six neutral residue types.

We calculate the weighted expected number of rotamers per atom
$\bar{f}$ for each residue group, where the weighting factor is taken
as the frequency of occurrence of the specific amino acid residue type
in eukaryotic proteins (see ref.\ \cite{TekaiaYeramianDujon02_G}). The
number of possible rotamers for each residue type is taken from
reference \cite{LovellRichardson00_P}.  The expected $\bar{f}$ values
are 1.10, 1.34, and 2.74 for hydrophobic, neutral, and polar amino
acid residues, respectively.  Polar residues have the largest
$\bar{f}$ value, but they are frequently involved in electrostatic
interactions and hydrogen bonding, which significantly decreases the
actual flexibility of polar side chains.  In general, $\bar{f}$ values
for natural amino acid residues are small, indicating that by the
criterion of weighted number of rotameric states per side chain
atom, they are rather inflexible.

It is remarkable that helix emerges as preferred main chain structure
for compact main chain conformations with maximum side chain entropy.
Our results indicate that the correlation between side chains plays
significant role in protein entropy and should be modeled more
accurately.  
\label{page:realProtein}
Real proteins have far more complex side chains.  For example, to
model a Lys residue realistically, a model of side chain of size 5
with all connecting flexibile bonds is needed.  The associated side
chain conformational space is much larger than the $M_3$ model
developed in this study, and therefore is not amenable to detailed
analysis.  However, we believe the conclusion obtained using $M_3$
model that inflexible side chain reduces folding entropy remains valid
if longer and more flexible side chain model is used.  In real
proteins, there are many residues whose side chains have flexibility
comparable to that of $M_3$ model ({\it e.g.}, His, Phe, Tyr, Val,
Ser, Cys, if we regard the inflexible part of their side chains as one
side chain bead in the $M_3$ model).  For these residues with reduced side chain flexibility, we find that
side chain entropy promotes the formation of helix for compact main
chain conformations.

Estimating side chain entropy is an important and difficult task for
modeling protein structure and protein stability.  With explicit side
chain models on three dimensional tetrahedral lattice, we have
developed an algorithm that calculates the exact side chain entropy of
tetrahedral lattice models for any given main chain structures of
moderate length. With current implementation, it works well up to
chain length of 30.  We compare results of side chain entropy
calculated by rotamer counting and by the exact method developed here.
\label{page:overcounting}
For longer chain polymers ({\it e.g.}, $N=30$), we found rotamer
counting method can give significantly over-estimated side chain
entropy.  For example, the average difference between the two methods
for models of length $N=30$ is larger for extended main chain
structures ($R_g >6.5$) and near compact main chain structures ($R_g =
2.7-2.9$, Figure~\ref{Fig:scEn_rg}d).

The method for exact calculation of side chain entropy given a
backbone structure can be generalized.  For longer chains, each
disconnected component in the conflict graph could contain too many
residues such that enumeration becomes infeasible. It is possible to
further develop an algorithm using the same divide-and-conquer
approach, where the large independent component is decomposed further
into two roughly equal size disconnected components by removing a
small number of edges in the conflict graph.  As an illustration, the
larger independent component in Figure~\ref{Fig:flx_alg} formed by
monomer 1, 3, 5, and 7 can be decomposed into two small disconnected
components by cutting the edge between vertices 4 and 6, which
corresponds to the shared position (labeled as ``a'').  The two
smaller components can then be enumerated separately, and side chains
of residues connected by the cut edges can also be enumerated
individually.  These enumeration will provide an exact value for the
total possible side chain arrangements of the original larger
independent component.  When a disconnected components contain a large
number of residues, an optimal decomposition becomes difficult.  This
is related to the graph partition problem. Although finding an optimal
solution to this problem is known to be an NP-complete problem
\cite{GareyJohnson79}, there are many effective approximation and
heuristic algorithms that are applicable for obtaining a good
decomposition.

Side chains in natural amino acid residues are chiral, and proteins
are better characterized using chiral models.  Chiral models developed
in this study will be useful for exploration of other properties of
proteins, where side chains play important roles.  Achiral models
introduced here may also be useful to study other polymers with
transient chirality on backbone \cite{HillMoore01_CHEMREV}, or
branched polymers such as peptoids, in which the chirality on nitrogen
atom is unstable and the side chain can easily convert between
opposite configuration \cite{KirshenbaumZuckermann98_PNAS}.

\section{Acknowledgment}
We thank Dr. Clare Woodward for helpful discussions.
This work is supported by grants from National Science Foundation
(CAREER DBI013356, DBI0078270, DMC0073601, CCR9980599), and National
Institute of Health (GM68958).  

\bibliography{chiral-flex2,pack,mathbio}
\bibliographystyle{unsrt}
\end{document}